\documentclass[aps,prl,twocolumn,superscriptaddress,appendix,showpacs,floatfix,nobibnotes]{revtex4}
\usepackage{epsfig}
\usepackage{epstopdf}

\usepackage{graphicx}
\usepackage{longtable}
\usepackage{CJK}
\usepackage{color}
\usepackage{mathptmx, courier, pifont}
\usepackage[scaled=0.92]{helvet}
\usepackage[T1]{fontenc}
\usepackage{textcomp}
\usepackage{soul}
\begin{document}

\begin{CJK*}{GB}{}

\title{Anomalous collective modes in atomic nuclei within the proton-neutron interacting boson model}

\author{Wei Teng}
\affiliation{Department of Physics, Liaoning Normal University,
Dalian 116029, P. R. China}

\author{Yu Zhang }\email{dlzhangyu_physics@163.com}
\affiliation{Department of Physics, Liaoning Normal University,
Dalian 116029, P. R. China}

\author{Sheng-Nan Wang}
\affiliation{Department of Physics, Liaoning Normal University,
Dalian 116029, P. R. China}

\author{Feng Pan}
\affiliation{Department of Physics, Liaoning Normal University,
Dalian 116029, P. R. China}\affiliation{Department of Physics and
Astronomy, Louisiana State University, Baton Rouge, LA 70803-4001,
USA}

\author{Chong Qi}
\affiliation{Department of Physics, KTH Royal Institute of
Technology, Stockholm 10691, Sweden}

\author{J. P. Draayer}
\affiliation{Department of Physics and Astronomy, Louisiana State
University, Baton Rouge, LA 70803-4001, USA}

\begin{abstract}
Novel collective modes characterized by a $B_{4/2}$ ratio ($\equiv B(E2;4_1^+\rightarrow 2_1^+)/B(E2;2_1^+\rightarrow 0_1^+)$) less than 1.0 that were observed recently have been identified within the proton-neutron interacting boson model (IBM-2) using the consistent-$Q$ Hamiltonian. These modes are shown to give rise to triaxial spectral features, including significant band mixing. The results provide a compelling explanation for the deeply suppressed $B_{4/2}$ ratio observed in $^{166}$W, $^{168,170}$Os, and $^{172}$Pt, offering new insights into the $B(E2)$ anomaly phenomenon in neutron-deficient nuclei.
\end{abstract}

\maketitle

\end{CJK*}

%\newpage
\begin{center}
\vskip.2cm\textbf{I. Introduction}
\end{center}\vskip.2cm

The emergence of collective features is one of the most important and striking characteristics of complex nuclear many-body systems. The interacting boson model (IBM) offers a suitable theoretical framework for addressing various nuclear collective modes using group or algebraic language~\cite{Iachellobook}. Typical collective modes may correspond to different dynamical symmetries in the IBM, such as U(5) (spherical vibrator), SU(3) (axially-deformed rotor), and O(6) ($\gamma$-unstable rotor). Despite the diversity in spectral structures among these modes, a common feature is the $B(E2)$ ratio $B_{4/2}\equiv B(E2;4_1^+\rightarrow2_1^+)/B(E2;2_1^+\rightarrow0_1^+)>1.0$, along with the energy ratio $R_{4/2}\equiv E(4_1^+)/E(2_1^+)\geq2$. These characteristics are consistent with various theoretical calculations and extensive experimental data on collective nuclei.
However, recent measurements on certain neutron-deficient nuclei near the neutron number $N_{\nu}$=90 and the proton dripline have challenged this established rule~\cite{Grahn2016,Saygi2017,Cederwall2018,Goasduff2019,Zhang2021}.
These measurements indicate an anomalous collective motion characterized by $R_{4/2}>2.0$ and $B_{4/2}\ll1.0$. This unusual phenomenon~\cite{Cakirli2004} has not been observed in conventional modes nor produced by calculations using microscopic approaches, posing a significant challenge to theoretical explanations~\cite{Grahn2016}. Recent studies~\cite{Zhang2022,Zhang2024,Pan2024,Wang2020,Zhang2025} using the phenomenological version of the interacting boson model (IBM-1) suggest that the $B(E2)$ anomaly, with $B_{4/2}<1.0$, can be explained by incorporating high-order terms~\cite{Heyde1984,Berghe1985,Vanthournout1988,Ramos2000I,Sorgunlu2008,Fortunato2011} to account for triaxial rotor dynamics~\cite{Leschber1987,Castanos1988,Smirnov2000,Zhang2014,Kotabook}.

Unlike IBM-1, which treats the degrees of freedom of protons and neutrons identically, the microscopic version of the IBM (the proton-neutron interacting boson model, or IBM-2)~\cite{Arima1977,Otsuka1978I} can naturally accommodate triaxiality-related collectivity~\cite{Dieperink1982,Arias2004,Caprio2004,Caprio2005} due to its explicit distinction between protons and neutrons. In recent years, the microscopic aspects of IBM-2~\cite{Otsuka1978II,Otsuka1981,Iachello1987} have been increasingly emphasized through a self-consistent derivation of the Hamiltonian based on microscopic mean-field calculations~\cite{Nomura2008,Nomura2010,Nomura2011I,Nomura2011II,Nomura2011III}. Consequently, IBM-2 is believed to provide a more comprehensive and quantitatively accurate framework for exploring various aspects of nuclear systems.

In this work, we investigate whether an unconventional mode with $B_{4/2}<1.0$ can be identified within the proton-neutron interacting boson model~\cite{Iachellobook}, thereby examining the previous rotor-mode explanation of the $B(E2)$ anomaly~\cite{Zhang2022,Zhang2024,Pan2024} in a more microscopic context.

\begin{center}
\vskip.2cm\textbf{II. Model Hamiltonian}
\end{center}\vskip.2cm

In the IBM, there are two types of bosons: the monopole $s$ boson with $J^\pi=0^+$
and the quadrupole $d$ boson with $J^\pi=2^+$. In IBM-2, the protons ($\pi$) and neutrons ($\nu$) bosons
are distinguished, and treated as bosons of two distinct types. The bilinear products of creation and annihilation operators for either proton or neutron bosons,
\begin{equation}
G_{\alpha,\beta}^{(\rho)}=b_{\rho,\alpha}^\dag b_{\rho,\beta},~~{(\alpha,\beta=1,\cdots,6)}\, ,\
\end{equation}
generate the U$_\rho$(6) ($\rho=\pi,~\nu$) group.
Accordingly, the proton-neutron coupled boson system
describe by the IBM-2 possesses the
direct product group symmetry,
$\mathrm{U}_\pi(6)\otimes \mathrm{U}_\nu(6)$,
describing a two-fluid system~\cite{Caprio2004,Caprio2005}.
The dynamical group $\mathrm{U}_\pi(6)\otimes \mathrm{U}_\nu(6)$
can be reduced to the rotational group $\mathrm{SO}_{\pi+\nu}(3)$
through various group chains associated with specific dynamical symmetry limits~\cite{Iachellobook}.
Physical operators and the Hamiltonian describing different symmetry limits and their mixing are constructed from the two types of boson operators, and the eigenvalue problem is solved by diagonalizing the Hamiltonian within a complete set of orthogonal bases corresponding to any one of the group chains~\cite{Iachellobook}. The widely used algorithm NPBOS~\cite{NPBOS} is to diagonalize the IBM-2 Hamiltonian in terms of the U(5) bases. Similarly, one can also solve the Hamiltonian using the SU(3) bases~\cite{Hu2023}.

In this work, we adopt the consistent-$Q$ Hamiltonian in the IBM-2, expressed as
\begin{eqnarray}\label{H}
\hat{H}=\varepsilon_0\Big[\eta(\hat{n}_{d_\pi}+\hat{n}_{d_\nu})-\frac{1-\eta}{N_\mathrm{B}}\hat{Q}_\pi^{\chi_\pi}\cdot\hat{Q}_{\nu}^{\chi_\nu}\Big]\, ,
\end{eqnarray}
where $\varepsilon_0>0$ and $\eta\in[0,1]$ are real parameters, $\hat{n}_{d_\rho}=d_\rho^\dag\cdot\tilde{d}_\rho$ is the $d$-boson number operator, and
\begin{eqnarray}\label{Q}
\hat{Q}_\rho^{\chi_\rho}=(s_\rho^\dag\times\tilde{d}_\rho+d_\rho^\dag\times\tilde{s}_\rho)^{(2)}+\chi_\rho(d_\rho^\dag\times\tilde{d}_\rho)^{(2)}\,
\end{eqnarray}
represents the quadrupole moment operator for protons ($\rho=\pi$) or neutrons ($\rho=\nu$).
Here, $N_\mathrm{B}=N_\pi+N_\nu$ denotes the total boson number, with  $N_{\pi(\nu)}$
being the number of valence nucleon or hole pairs counted from the nearest shells for a given nucleus.
The consistent-Q Hamiltonian in (\ref{H}) captures essential features of low-lying quadrupole collective states~\cite{Talmibook}. For instance, this simple Hamiltonian form has been used to illustrate the spectral properties of neutron-rich Os and Pt isotopes for the mass number A$\sim$180{--}200~\cite{Nomura2011II,Nomura2011III}, and its IBM-1 counterpart has been extensively employed in discussions of shape phase transitions in nuclei\cite{Cejnar2009}.

The $E2$
transition operator is simply adopted as
\begin{eqnarray}\label{TE2}
\hat{T}(E2)=e_\pi\,\hat{Q}_\pi^{\chi_\pi}+e_\nu\,\hat{Q}_\nu^{\chi_\nu}\, ,
\end{eqnarray}
where
$e_{\rho}$ with $\rho=\pi,\nu$ denotes the
effective charge parameter
and $\hat{Q}_{\rho}^{\chi_{\rho}}$
is taken the same as
in the Hamiltonian (\ref{H}).
The reduced $E2$ transition
rate is calculated via
\begin{eqnarray}
B(E2;L_i^+\rightarrow L_f^+)=\frac{|\langle\alpha_fL_f^+\parallel \hat{T}(E2)\parallel\alpha_iL_i^+\rangle|^2}{2L_i+1}\, ,
\end{eqnarray}
where
$\alpha$ represents
the quantum numbers
other than the angular momentum $L$ for a given state.

\begin{center}
\vskip.2cm\textbf{III. Results and Discussions}
\end{center}\vskip.2cm

According to previous analyses based on IBM-1~\cite{Zhang2022,Zhang2024,Zhang2025}, the results of  $B_{4/2}<1.0$ can be produced from finite-$N$ triaxial rotor modes established through the boson mapping of a triaxial rotor~\cite{Zhang2014}, necessitating the inclusion of high-order interaction terms in the IBM-1 Hamiltonian. In contrast, triaxial features in IBM-2~\cite{Dieperink1982} can be generated by considering only quadrupole-quadrupole interactions, where the quadrupole operators for neutrons and protons take different forms, such as setting $\chi_\nu=-\chi_{\pi}$ in Eq.~(\ref{Q}). As noted in~\cite{Caprio2005}, the microscopic conditions that lead to a two-fluid triaxial structure, characterized by $\chi_\pi$ and $\chi_{\nu}$
of opposite signs, are found when proton bosons are particle-like (below mid-shell) and neutron bosons are hole-like (above mid-shell), or vice versa. It should be mentioned that, to achieve the ideal triaxial limit (SU(3)$_{\pi\nu}^\ast$ symmetry)~\cite{Dieperink1982} in IBM-2, one needs in principle to include as well the quadrupole-quadrupole interaction of like bosons $\hat{Q}_\rho\cdot\hat{Q}_\rho$ $(\rho=\pi,~\nu)$. However, in practice, the proton-neutron quadrupole interaction dominates over the proton-proton and neutron-neutron interactions~\cite{Nomura2011II,Nomura2011III}. Therefore, for simplicity, we will utilize the consistent-Q Hamiltonian defined in (\ref{H}) for our theoretical analyses throughout this work. In the following sections, we will first identify the specific model conditions for producing a collective mode with  $B_{4/2}<1.0$ and then examine the IBM-2 explanation of the  $B(E2)$ anomaly observed in experiments.

\begin{center}
\vskip.2cm\textbf{(A). Numerical examinations}
\end{center}\vskip.2cm

\begin{figure*}
\begin{center}
\includegraphics[scale=0.24]{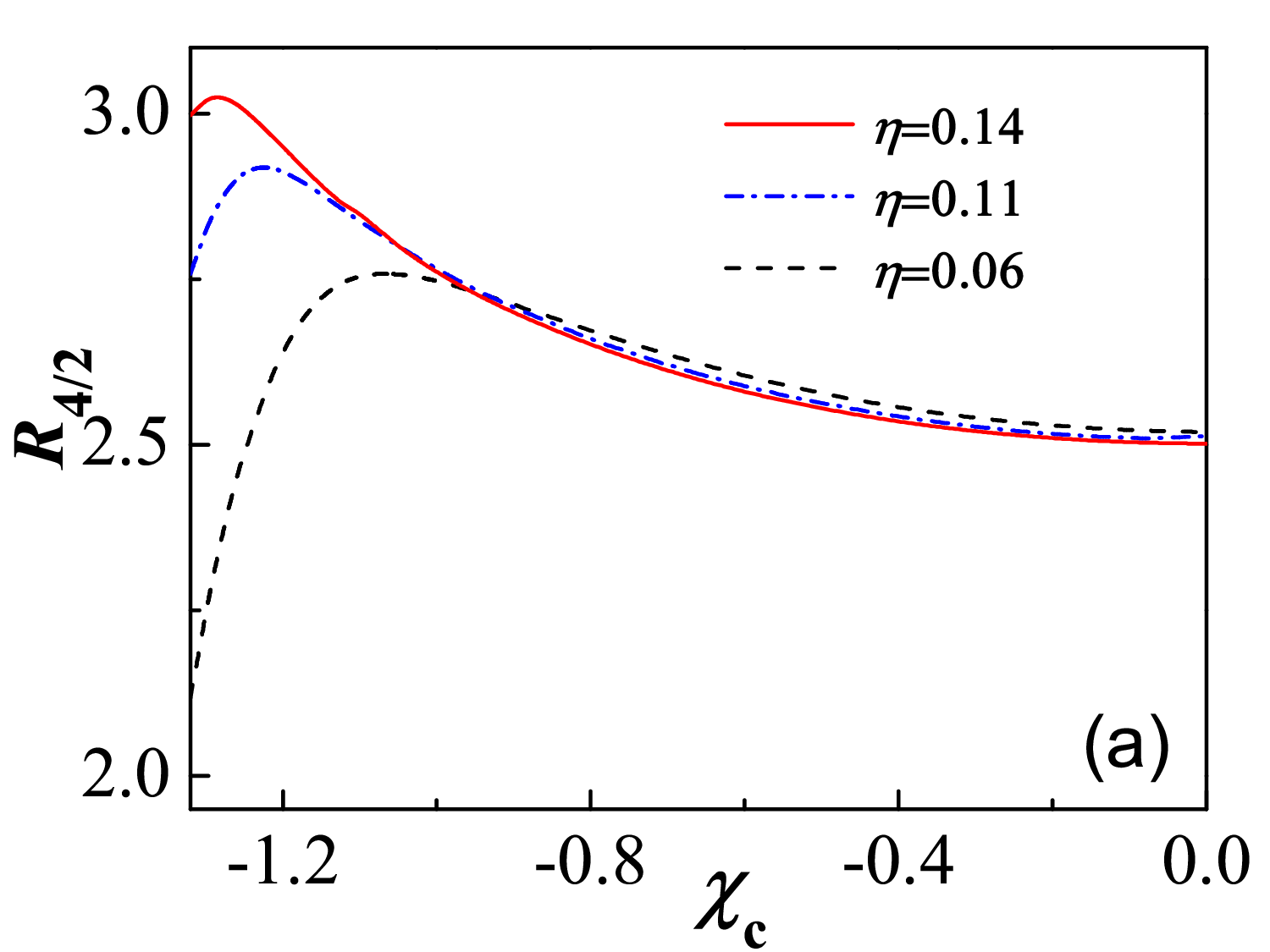}
\includegraphics[scale=0.24]{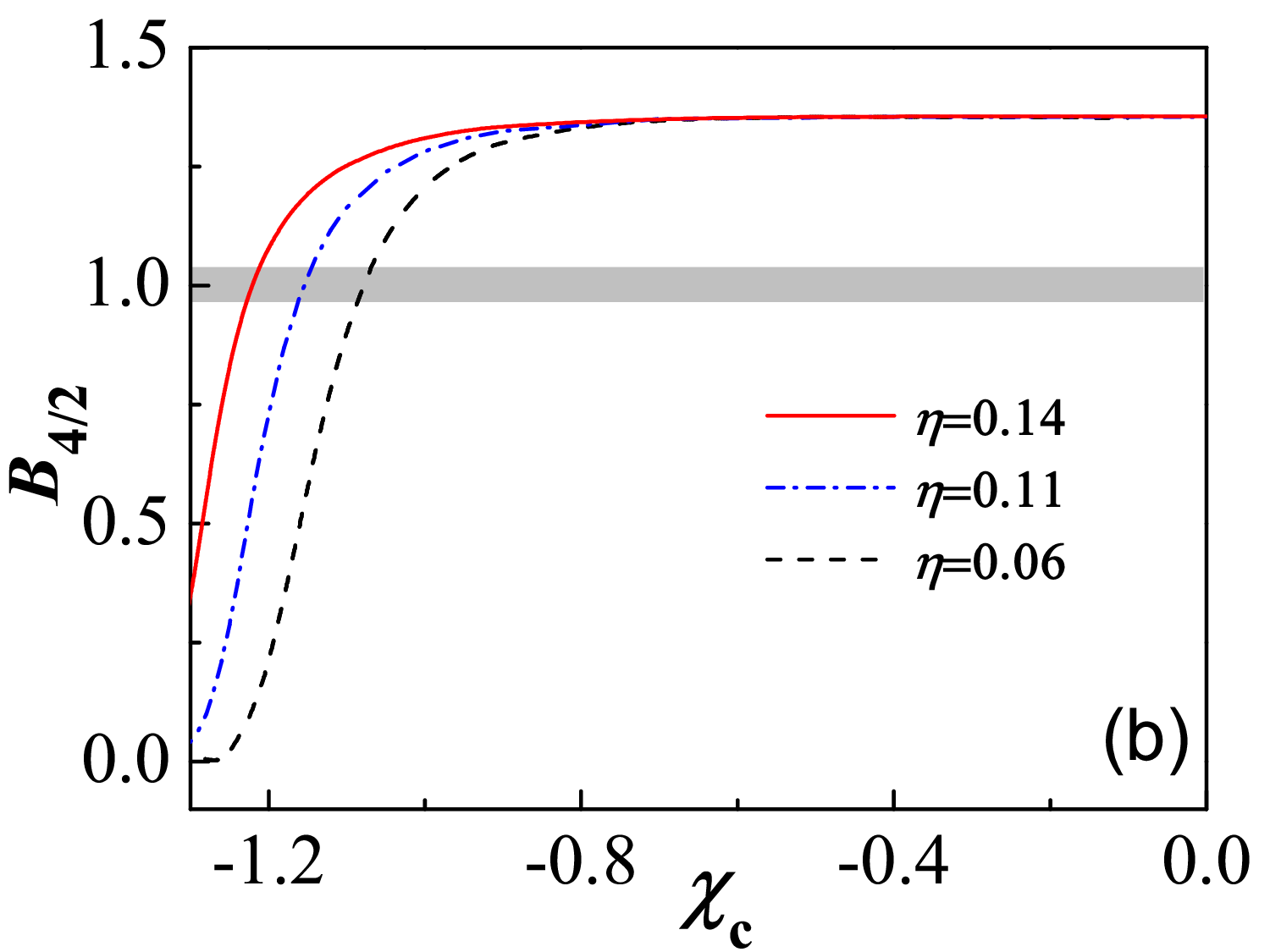}
\includegraphics[scale=0.24]{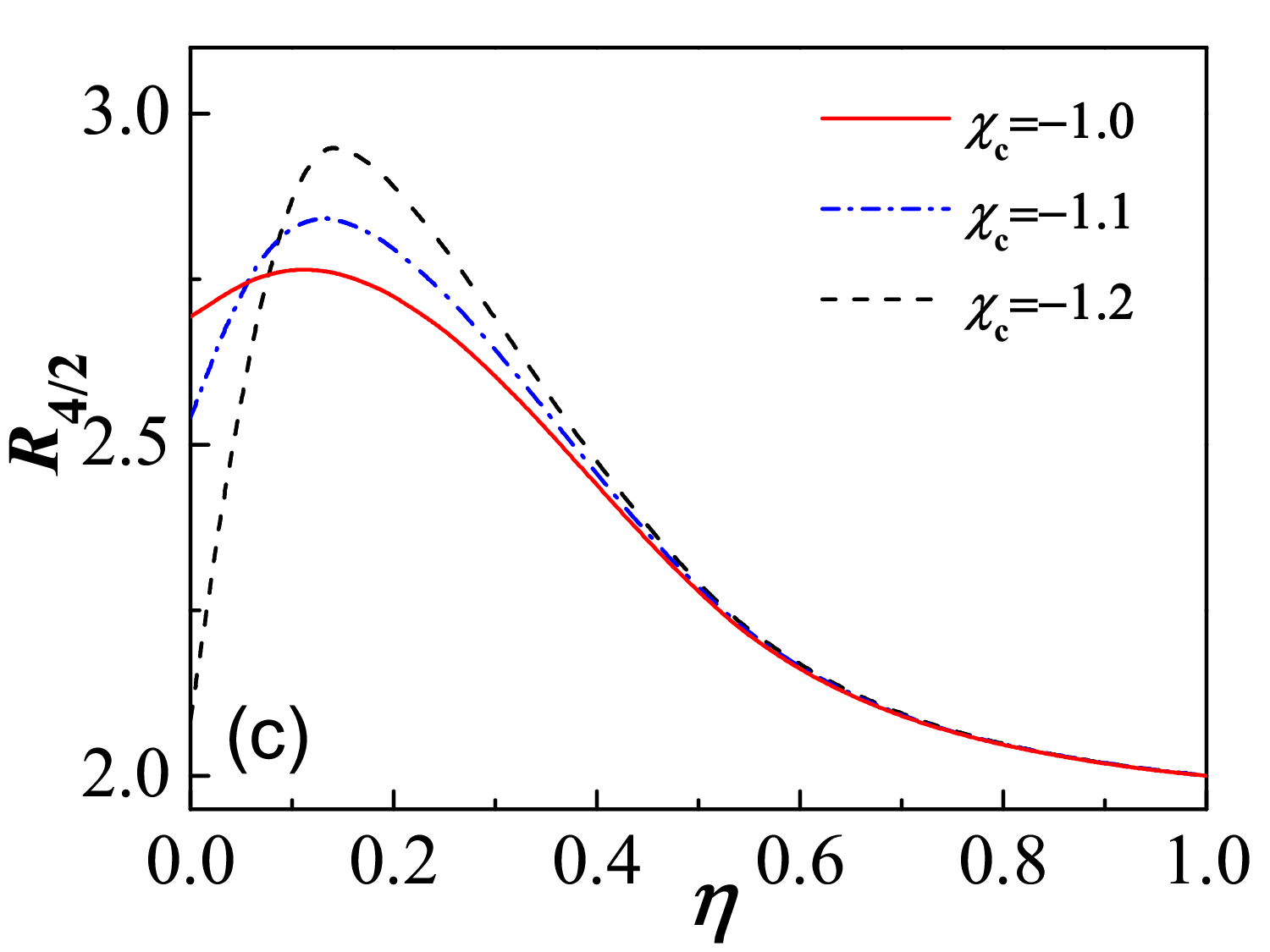}
\includegraphics[scale=0.24]{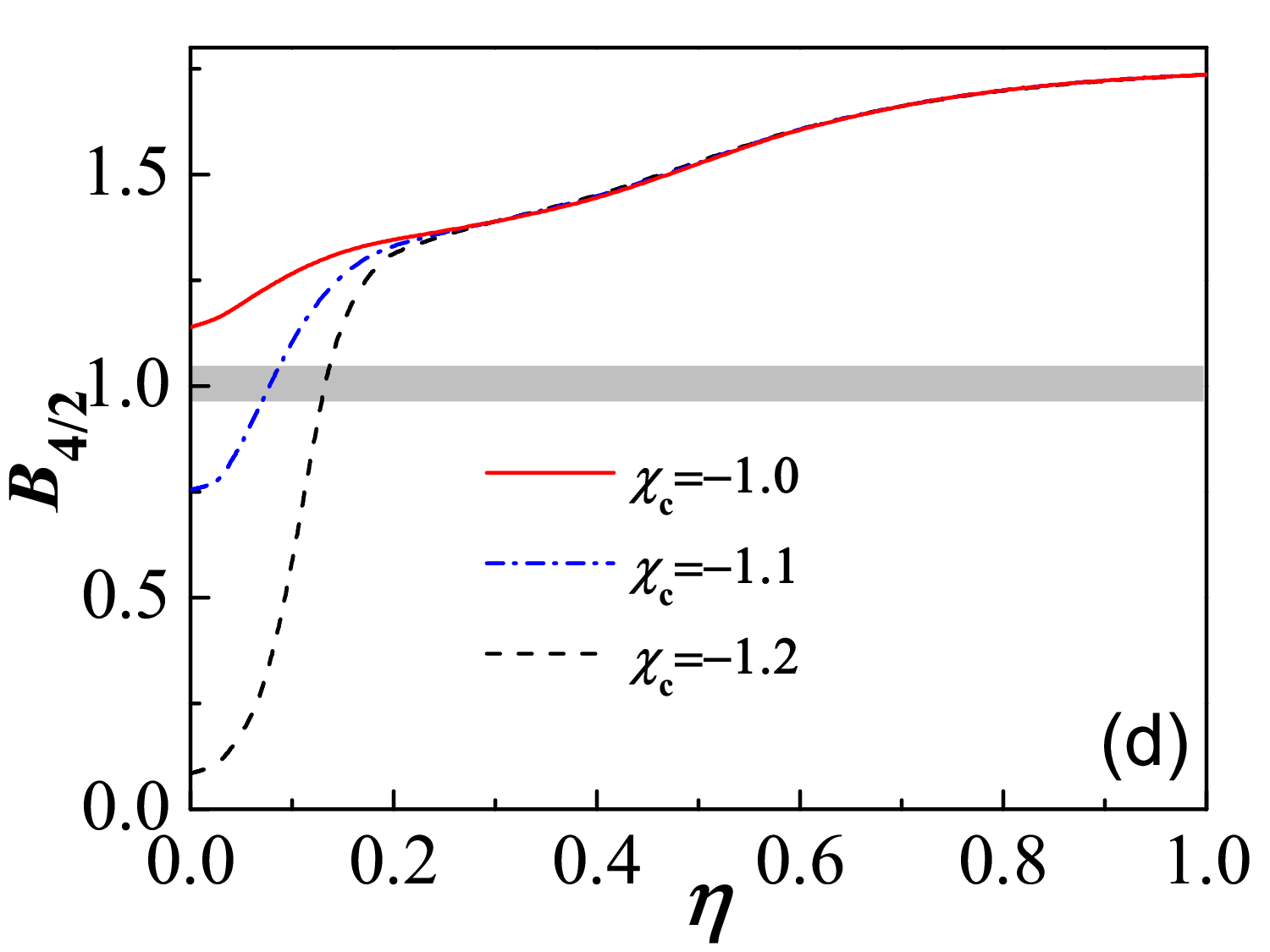}
\caption{(a) The results for the $R_{4/2}$ ratio derived from the consistent-$Q$ Hamiltonian ($N_\pi=3$ and $N_\nu=5$) for different $\eta$ are
presented as a function of $\chi_\mathrm{c}$ (see the tex for its definition). (b) The same as in (a) but for the $B(E2)$ ratio $B_{4/2}$. (c) The results for the $R_{4/2}$ ratio derived from the Hamiltonian for
different $\chi_\mathrm{c}$ are
presented as a function of $\eta$. (d) The same as in (c) but for the $B(E2)$ ratio $B_{4/2}$.} \label{F1}
\end{center}
\end{figure*}

To investigate the influences of different parameters on collective features, the ratios $R_{4/2}$ and $B_{4/2}$ are
calculated from the Hamiltonian (\ref{H})
by setting $\chi_\nu=-\chi_\pi=\chi_\mathrm{c}$ for convenience without loosing generality.
To simulate a realistic system associated with $B(E2)$
anomaly~\cite{Grahn2016},
the boson numbers in the calculations are chosen as $N_\pi=3$
and $N_\nu=5$, corresponding to $^{168}$Os.
Since $\pm\chi_\mathrm{c}$ yield the same results,
only $\chi_\mathrm{c}\leq0$ is
considered in the discussions.
Furthermore, for the $B(E2)$ calculations,
it is assumed that the effective charge parameters
are equal, $e_\pi=e_\nu$, for simplicity, as
often adopted in the IBM-2 calculations~\cite{Iachellobook,Nomura2010,Nomura2011III,Isacker1980}.
The calculated results are presented as a function of either $\chi_\mathrm{c}$ or $\eta$ in Fig.~\ref{F1}.
As observed in Fig.~\ref{F1}(a),
the $R_{4/2}$ ratio as a function of $\chi_\mathrm{c}$
exhibits a non-monotonic evolution,
but its values  remain above
$R_{4/2}>2.0$, as expected for a collective mode.
In contrast, the results in Fig.~\ref{F1}(b) shows
that the $B_{4/2}$ ratio drops
to significantly low values,
$B_{4/2}\ll1.0$ when $\chi_\mathrm{c}<-1.1$,
and then quickly stabilizes at around $B_{4/2}=1.35$ within $\chi_\mathrm{c}\in[-0.8,~0]$.
This striking feature demonstrates that unconventional $E2$ structures
emerge from the coupled two-fluid systems
dominated by quadruple-quadrupole interaction.
Adopting $\chi_\pi=-\chi_\nu$
suggests that such exotic features are more likely when proton and neutron bosons have different quadrupole configurations and, thus, different quadrupole moment distributions. Mean-field analyses of the IBM-2 phase diagram in~\cite{Arias2004, Caprio2004, Caprio2005} indicate that proton and neutron bosons in coupled systems may exhibit different quadrupole deformations when
$\chi_\pi=-\chi_\nu$.
Conversely,
the consistent-$Q$
Hamiltonian with $\chi_\pi\simeq\chi_\nu$ may
produce results for low-lying states
similar to those obtained from the IBM-1 counterpart~\cite{Iachello2004,Giannatiempo2012}.
However, numerical calculations show
that no $B_{4/2}<1.0$ result is
achieved if $\chi_\pi$ and $\chi_\nu$ have the same sign.
As further shown in panels (c) and (d),
the results indicate that the ratio $B_{4/2}$ may rapidly increase
from $B_{4/2}<1.0$ to $B_{4/2}>1.0$ by continuously increasing
the weight of the U(5) component ($\eta$).
Meanwhile, the energy ratio $R_{4/2}$
as a function of $\eta$ exhibits
non-monotonic evolution and converges
towards the vibrational limit ($R_{4/2}\simeq2.0$) at
large $\eta$. Interestingly, $R_{4/2}$ in these cases may reach
its maximal values around $\eta=0.2$, roughly coinciding
with $B_{4/2}\simeq1.0$, This suggests a phase transition-like evolution between
the $E2$-anomalous modes with $B_{4/2}<1.0$ and those normal modes
with $B_{4/2}>1.0$. Note that this transitional behavior is not assumed to be directly associated with the spherical to deformed quantum phase transitions described by
the consistent-$Q$ Hamiltonian (\ref{H}) under $\chi_\pi=-\chi_\nu$, which are anticipated to occur around $\eta\approx0.5$ in the large-$N$ limit.

\begin{figure}
\begin{center}
\includegraphics[scale=0.17]{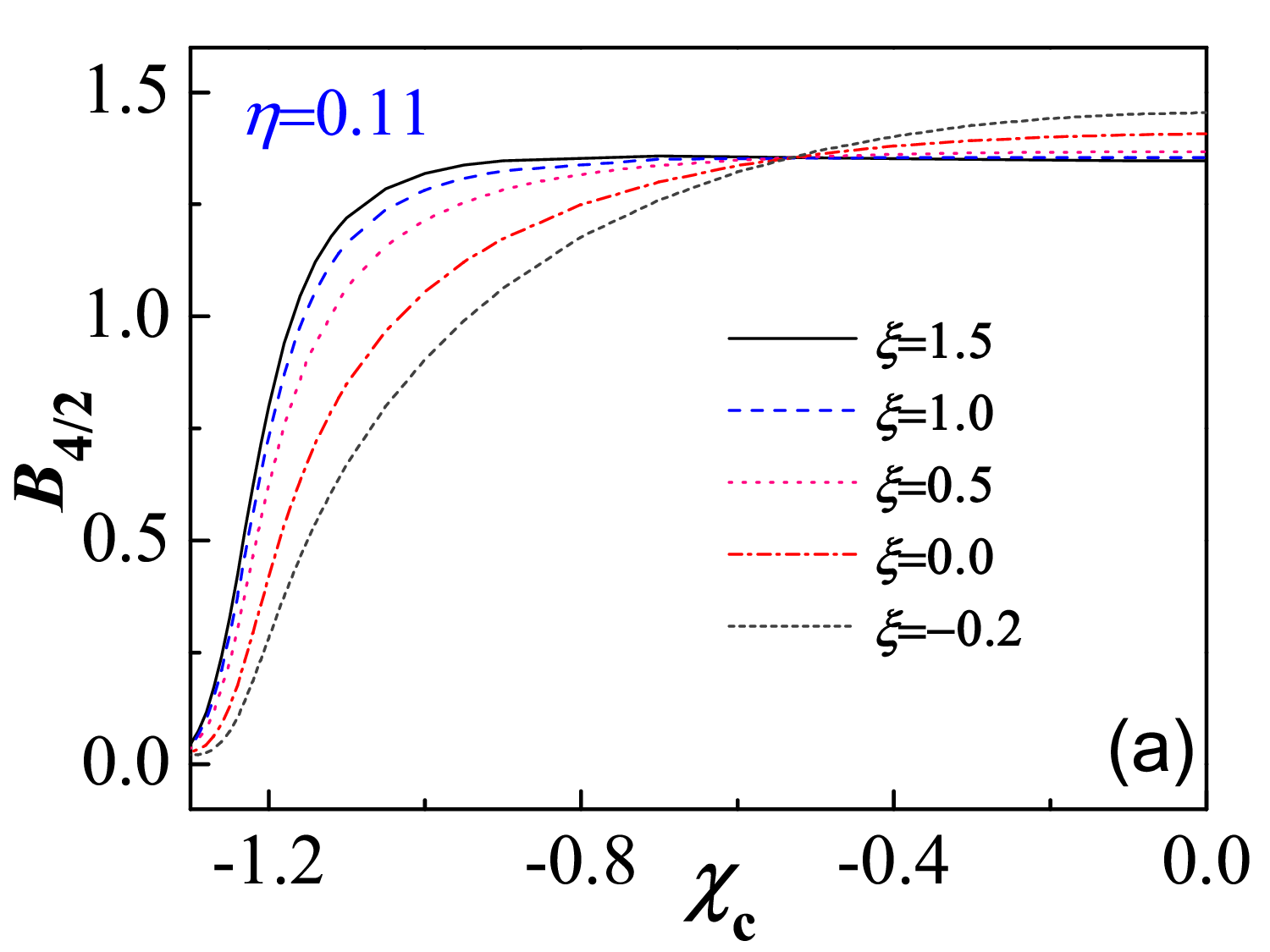}~\includegraphics[scale=0.17]{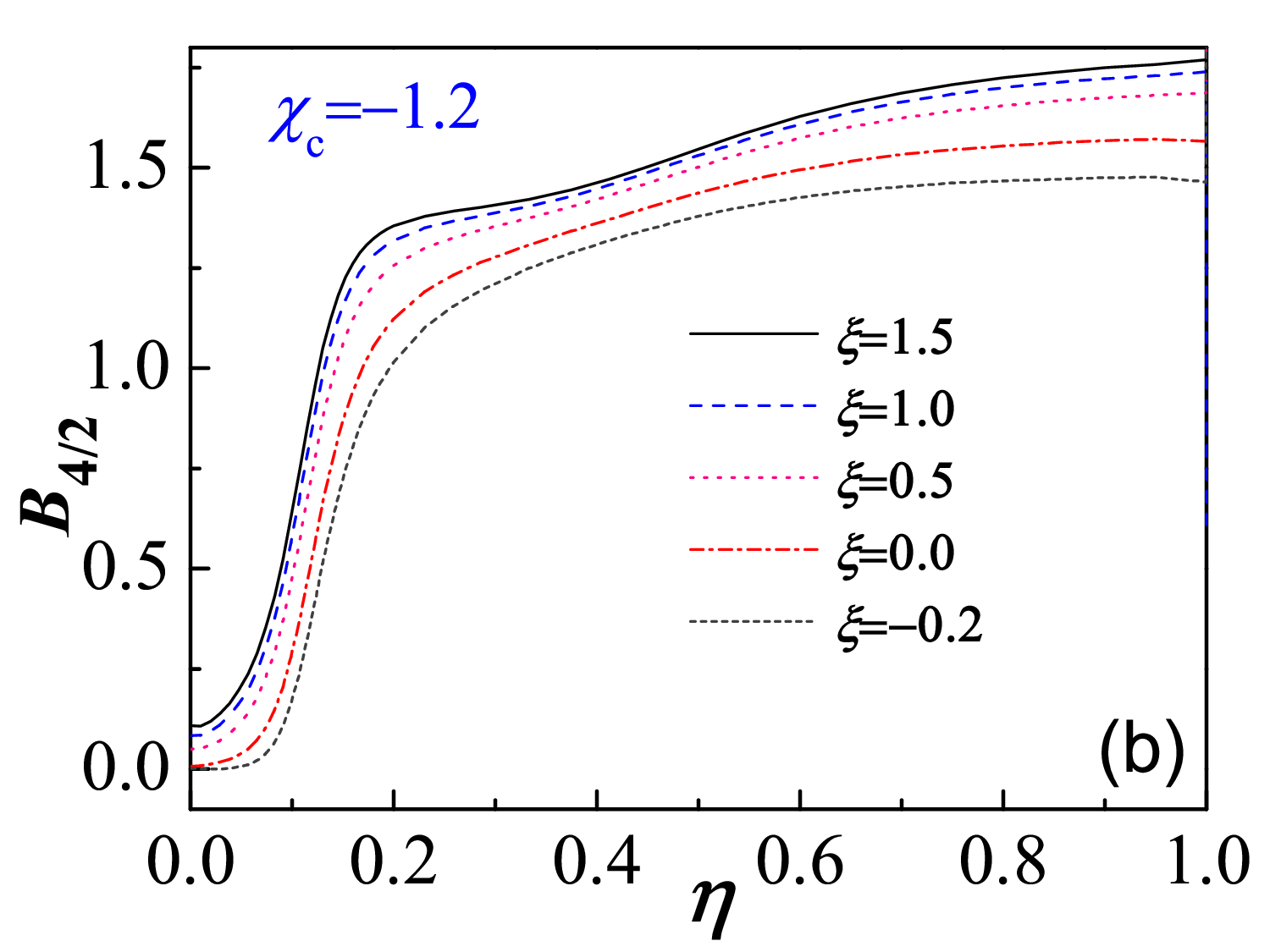}
\includegraphics[scale=0.17]{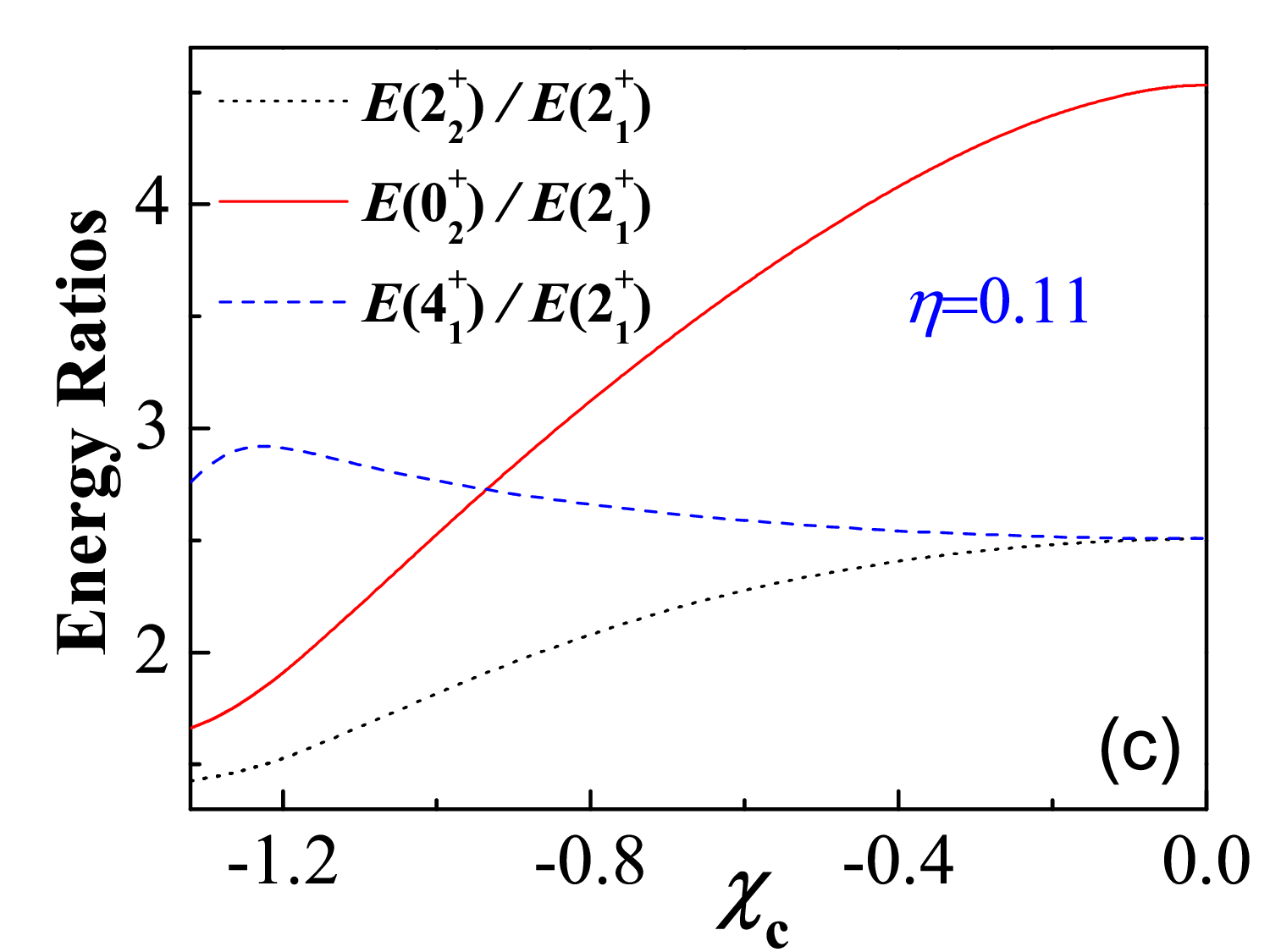}~\includegraphics[scale=0.17]{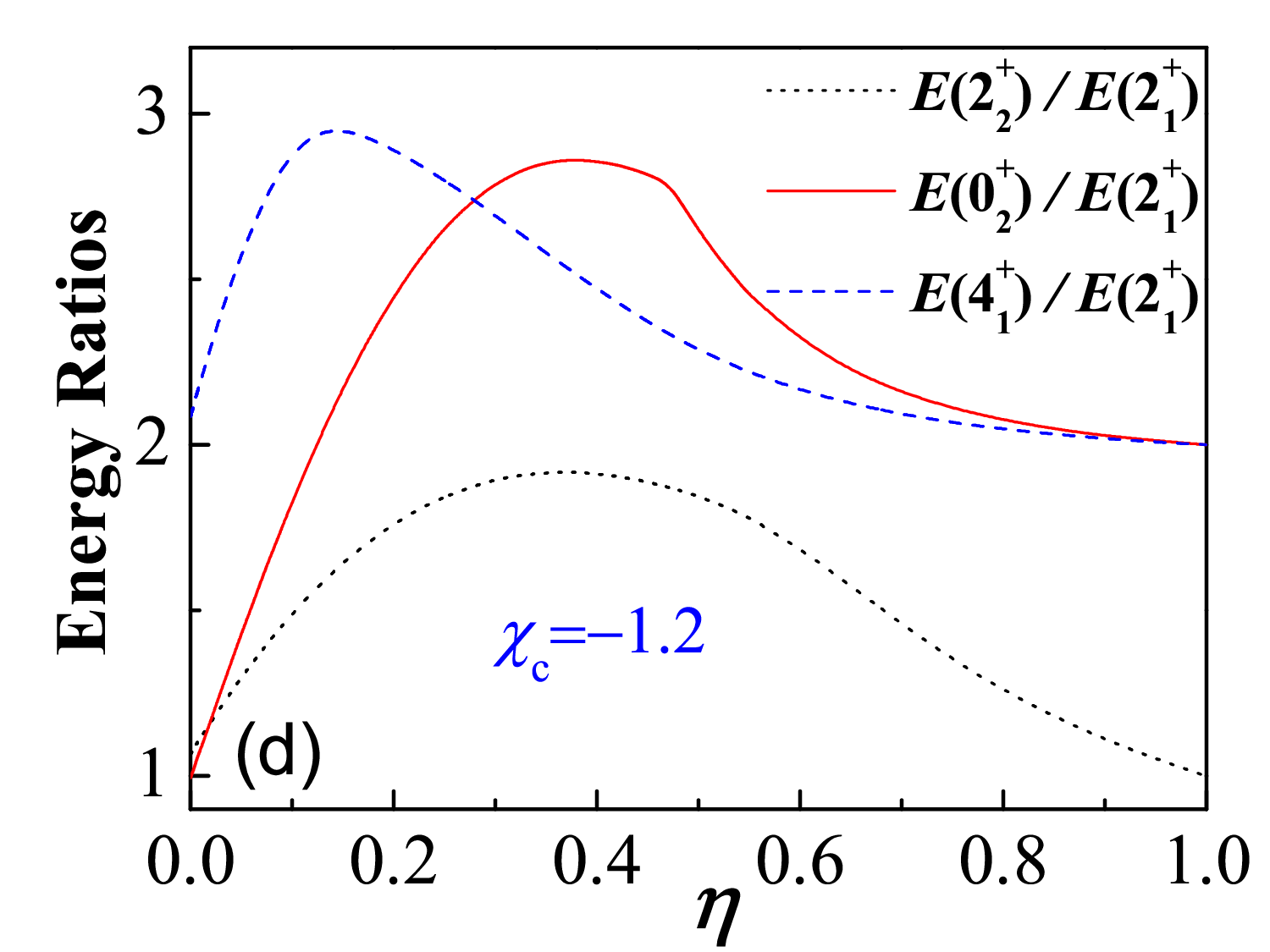}
\caption{(a) Evolution of the $B_{4/2}$ ratio solved from the case with $\eta=0.11$ given in Fig.~\ref{F1}(b) but for different effective charges. (b) The same as in (a) but for the case with $\chi_\mathrm{c}=-1.2$ derived from Fig.~\ref{F1}(d). (c) Evolution of typical energy ratios solved from the case (a). (d) Evolution of typical energy ratios solved from the case (b).
}\label{F12}
\end{center}
\end{figure}

Since the adopting
different effective charges $e_{\pi(\nu)}$
in the $E2$ transition operator can
influence the $B(E2)$ results to some extent,
we examine two specific cases:
one with $\eta=0.11$ (as shown in Fig.~\ref{F1}(b))
and another with $\chi_\mathrm{c}=-1.2$ (as shown in Fig.~\ref{F1}(d)). The calculated $B_{4/2}$ results for different ratios of $\xi=e_\nu/e_\pi$,
along with typical energy ratios are presented in Fig.~\ref{F12}.
As observed in Fig.~\ref{F12}(a) and Fig.~\ref{F12}(b), varying $\xi$ does not qualitatively alter the evolutional characters of the $B_{4/2}$ ratio as a function of either $\chi_\mathrm{c}$ or $\eta$.
This indicates that the observed $B(E2)$ anomaly is primarily determined by the eigenstates.
Therefore, unless otherwise specified, we will continue to
use $e_\pi=e_\nu$ in the subsequent discussions. It should be noted that the effective charges with $e_\nu/e_\pi\leq0$ in Fig.~\ref{F12} are included solely for the purpose of comparison analysis and never utilized in actual calculation~\cite{Isacker1980,Ginocchio1986,Wolf1992}. Additionally, the results suggest
that low-energy non-yrast states of $0_2^+$ and $2_2^+$ may
constantly appear in cases where $B_{4/2}<1.0$, as seen from Fig.~\ref{F12}(c) and Fig.~\ref{F12}(d).
These features are consistent with triaxiality, which is typically characterized by a low-lying $K=2$ band. This point will be further discussed below.

\begin{figure}
\begin{center}
\includegraphics[scale=0.24]{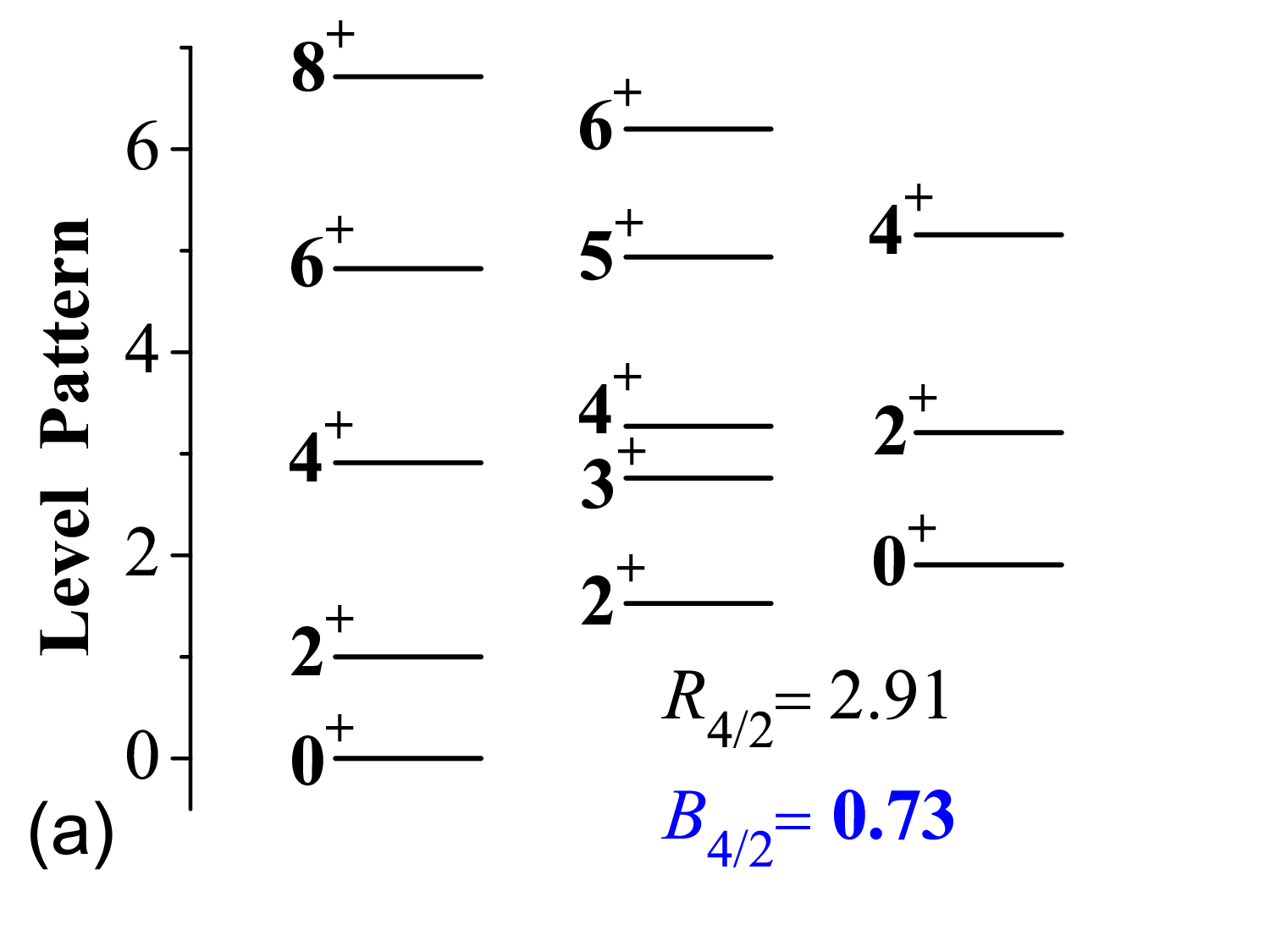}
\includegraphics[scale=0.24]{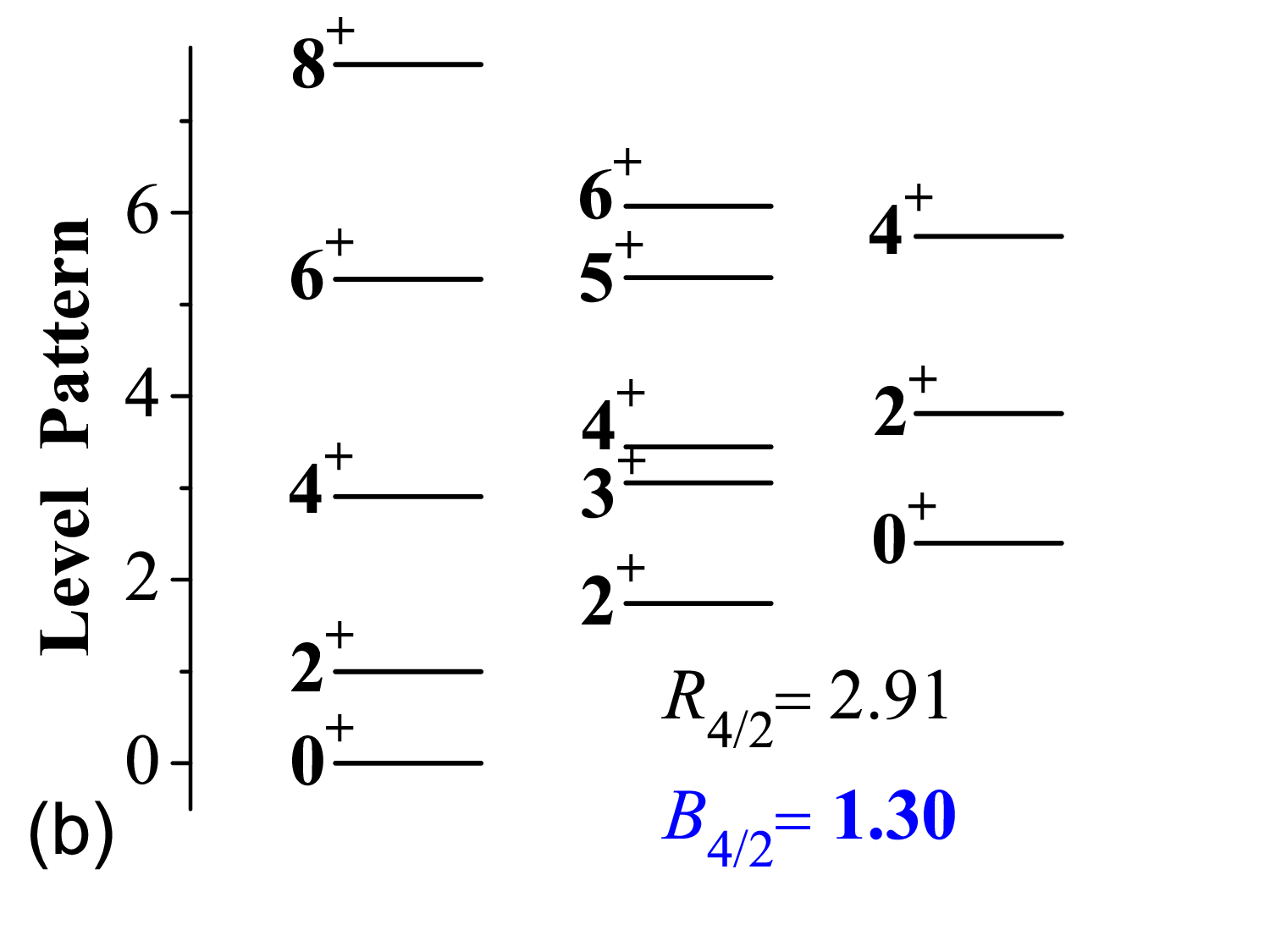}
\caption{(a) The low-lying level pattern derived from the consistent-$Q$ Hamiltonian ($N_\pi=3$ and $N_\nu=5$) with $\eta=0.11$ and $\chi_\nu=-\chi_\pi=-1.2$. (b) The same as in (a) but for that derived from the Hamiltonian with $\eta=0.19$.}\label{F2}
\end{center}
\end{figure}

To exemplify the spectral characteristics of an unconventional mode with $B_{4/2}<1.0$,
level pattern solved from the consistent-$Q$ Hamiltonian with $\eta=0.11$ and $\chi_\nu=-\chi_\pi=-1.2$ is
presented in Fig.~\ref{F2}(a).
For comparison, the case for $B_{4/2}>1.0$ obtained by increasing the U(5) weight ($\eta=0.19$) in the Hamiltonian is provided in Fig.~\ref{F2}(b).
As observe from Fig.~\ref{F2}(a),
a $B(E2)$ anomaly structure with $B_{4/2}=0.73$ and $R_{4/2}=2.91$ has indeed emerged in the IBM-2 system,
accompanied by the low-energy $\gamma$ and $\beta$ bands.
The appearance of excited rotational bands at low energy
is typically a sign of a triaxial system
and is suggested
to be associated with band mixing
effects~\cite{Wood2004,Allmond2008}.
This point can be partially inferred
from the $B(E2)$ ratios for the relevant states, such as those with $L^\pi=0^+,~2^+,~4^+,~\cdots$ displayed in Fig.~\ref{F2}. Apart from  $B_{4/2}=0.73$, the $B(E2)$ anomaly system may yield the following ratio values: $\frac{B(E2;4_1^+\rightarrow2_2^+)}{B(E2;2_1^+\rightarrow0_1^+)}=0.49$, $\frac{B(E2;4_1^+\rightarrow2_3^+)}{B(E2;2_1^+\rightarrow0_1^+)}=0.08$, $\frac{B(E2;4_2^+\rightarrow2_1^+)}{B(E2;2_1^+\rightarrow0_1^+)}=0.60$, $\frac{B(E2;4_2^+\rightarrow2_2^+)}{B(E2;2_1^+\rightarrow0_1^+)}=0.32$ and $\frac{B(E2;4_2^+\rightarrow2_3^+)}{B(E2;2_1^+\rightarrow0_1^+)}=0.03$.
These results clearly demonstrate that there is a strong mixing between yrast band (ground-state band) and yrare band ($\gamma$ band) in the present case, suggesting that the anomalous depressed $B_{4/2}$ value observed in Fig.~\ref{F2}(a) is primarily caused by the band-mixing effects.
As further observed from Fig.~\ref{F2}(b), a rather similar level pattern with $R_{4/2}=2.91$ has been generated even by increasing the weight of the U(5) term in the Hamiltonian (\ref{H}),
which, however, yields a "normal" $B(E2)$ result with $B_{4/2}=1.30$. This observation is consistent with the different monotonicity in the evolutions of $R_{4/2}$ and $B_{4/2}$ as depicted in Fig.~\ref{F1}.
In this case, the relevant $B(E2)$ ratios are $\frac{B(E2;4_1^+\rightarrow2_2^+)}{B(E2;2_1^+\rightarrow0_1^+)}=0.11$, $\frac{B(E2;4_1^+\rightarrow2_3^+)}{B(E2;2_1^+\rightarrow0_1^+)}=0.02$, $\frac{B(E2;4_2^+\rightarrow2_1^+)}{B(E2;2_1^+\rightarrow0_1^+)}=0.05$, $\frac{B(E2;4_2^+\rightarrow2_2^+)}{B(E2;2_1^+\rightarrow0_1^+)}=0.76$ and $\frac{B(E2;4_2^+\rightarrow2_3^+)}{B(E2;2_1^+\rightarrow0_1^+)}=0.10$.
Clearly, the inter-band $E2$ transitions between yrast and yrare states become much weaker compared to the corresponding ones in the $B_{4/2}<1.0$ case discussed earlier.
This suggests that
increasing the contribution from the U(5) term enhances the intra-band $E2$ transitional strengths by reducing the band mixing
effects, ultimately leading to a $B_{4/2}>1.0$ result as expected in conventional collective modes.

\begin{center}
\vskip.2cm\textbf{(B). Mean-field analysis}
\end{center}\vskip.2cm

To comprehend these exotic modes characterized by $R_{4/2}<1.0$ at the mean-field level,
one can examine the potential energy per boson using
the coherent state method~\cite{Iachellobook}.
The coherent %(instate)
state for IBM-2 is defined as
\begin{eqnarray}\label{coherent}
|N_\pi,N_\nu,\beta_\pi,\beta_\nu,\gamma_\pi,\gamma_\nu\rangle=\sqrt{\frac{1}{N_\pi!N_\nu!}}(B_\pi^\dag)^{N_\pi}(B_\nu^\dag)^{N_\nu}|0\rangle\,
\end{eqnarray}
with
\begin{eqnarray}\nonumber
&B_\rho^\dag=\sqrt{\frac{1}{1+\beta_\rho^2}}\Big[s_\rho^\dag+\beta_\rho \mathrm{cos}(\gamma_\rho)d_{\rho,0}^\dag+\\
&~~~~~~~~~~~~~~~~~~~~~~~~~~~~~~~~~~~~\frac{1}{\sqrt{2}}\beta_\rho\mathrm{sin}(\gamma_\rho)\Big(d_{\rho,2}^\dag+d_{\rho,-2}^\dag\Big)\Big]\, .
\end{eqnarray}
It should be noted that, according to previous mean-field analyses of the IBM-2~\cite{Arias2004,Caprio2005,Ginocchio1992}, the relative Euler angles between the proton and neutron fluids are assumed to be zero in the present context. The potential function is then defined as the expectation value of the interaction within the coherent state (\ref{coherent}). Specifically, the potentials for two parts involved in the Hamiltonian (\ref{H}) can be derived as~\cite{Arias2004,Caprio2004,Caprio2005,Giannatiempo2012}
\begin{eqnarray}\label{V1}
V_{n_d}&\equiv&\langle\hat{n}_{d_\pi}+\hat{n}_{d_\nu}\rangle=\frac{N_\pi \beta_\pi^2}{1+\beta_\pi^2}+\frac{N_\nu \beta_\nu^2}{1+\beta_\nu^2}\, ,
\end{eqnarray}
and
\begin{eqnarray}\label{V}
V_{QQ}&\equiv&\langle-\hat{Q}_\pi\cdot\hat{Q}_\nu\rangle\\ \nonumber
&=&-\frac{2N_\pi N_\nu\beta_\pi\beta_\nu}{7(1+\beta_\pi^2)(1+\beta_\nu^2)}\\ \nonumber
&\times&\Big[14\mathrm{cos}(\gamma_\pi-\gamma_\nu)+\chi_\pi\chi_\nu\beta_\pi\beta_\nu \mathrm{cos}(2\gamma_\pi-2\gamma_\nu)\\ \nonumber
&-&\sqrt{14}\Big(\chi_\pi\beta_\pi\mathrm{ cos}(2\gamma_\pi+\gamma_\nu)+\chi_\nu\beta_\nu \mathrm{cos}(2\gamma_\nu+\gamma_\pi)\Big)\Big]\, .
\end{eqnarray}

The potential forms given in (\ref{V1}) and (\ref{V}) are precisely identical to those previously derived~\cite{Arias2004,Caprio2004,Caprio2005,Giannatiempo2012} using the same coherent state definition~\cite{Iachellobook}.
It is clear that the mean-field structure of the IBM-2 is more intricate compared to the IBM-1~\cite{VC1981},
as the ground-state geometry in the IBM-2 is governed by four deformation variables, $\beta_\rho$ and $\gamma_\rho$ with $\rho=\pi,\nu$.
In contrast to the relatively simple potential structure of the single-body term $V_{n_d}$ in (\ref{V1}), the two-body term $V_{QQ}$ given in (\ref{V}) exhibit
a highly complex dependence on $\beta_\rho$ and $\gamma_\rho$. Nonetheless, the classical potential depends on boson number solely in a scaling way, implying that quadrupole deformation extracted from the two-body term $V_{QQ}$ can be expressed as an analytical function of $\chi_\mathrm{c}$, independent of $N_\pi$ and $N_\nu$. However, the mean-field results become explicitly $N$-dependent once the single-body term $V_{n_d}$ is included. For simplicity, we will focus here on a mean-field analysis of the two-body term given in (\ref{V}) and leave the mean-field discussion of the full Hamiltonian involving more numerical calculations elsewhere. This is mainly because $B(E2)$ anomaly in the current model is primarily induced by the quadrupole-quadrupole interaction, as illustrated in Fig.~\ref{F1}. Besides, this may enable us to identify the mean-field deformation analytically, as demonstrated below. In the following, we will concentrate on comparing two scenarios described by the two-body potential $V_{QQ}$ in (\ref{V}) with $\chi_\nu=\mp\chi_\pi$. More general mean-field analysis regarding the role of different interactional terms in the IBM-2 can be found in \cite{Caprio2005}.

For the cases where
$\chi_\nu=-\chi_\pi=\chi_\mathrm{c}$, which are
favoured by the $B(E2)$ anomaly,
a $\chi$-independent constant for the ground-state energy,
\begin{eqnarray}\label{Eg}
E_g\equiv{V_{QQ}}_{\mathrm{min}}(\bar{\beta}_\pi,\bar{\beta}_\nu,\bar{\gamma}_\pi,\bar{\gamma}_\nu)|_{-\chi_\pi=\chi_\nu=\chi_\mathrm{c}}=-N_\pi N_\nu\,
\end{eqnarray}
can be derived by
minimizing the potential $V_{QQ}$ with respect to the deformation parameters $(\beta_\rho,\gamma_\rho)$.
This minimization results in
two sets of degenerate optimal values,
which can be analytically expressed as $(\bar{\beta}_\pi,\bar{\beta}_\nu,\bar{\gamma}_\pi,\bar{\gamma}_\nu)=(A-B,~A+B,~0^\circ,~0^\circ)$ and $(A+B,~A-B,~60^\circ,~60^\circ)$,
with $A=\sqrt{14+\chi_\mathrm{c}^2}/\sqrt{14}$ and $B=\chi_\mathrm{c}/\sqrt{14}$. It is noteworthy that the optimal parameters consistently satisfy $\bar{\beta}_\pi\bar{\beta}_\nu=1$ for any given $\chi_\mathrm{c}$ values.
Consequently, the ground state energy expressed in (\ref{Eg}) can be rewritten as $E_g=-N_\pi N_\nu\bar{\beta}_\pi\bar{\beta}_\nu$, which is consistent with the mean-field expressions (Eq.(4.6) with $\kappa_{\pi\pi}^\prime=\kappa_{\nu\nu}^\prime=0$) derived in \cite{Caprio2005}. Evidently, these two degenerate minima represent two distinct $\gamma$ deformations:
($\bar{\gamma}_\pi=\bar{\gamma}_\nu=0^\circ$) and ($\bar{\gamma}_\pi=\bar{\gamma}_\nu=60^\circ$).
Apart from that, proton and neutron bosons in these cases may also exhibit different $\beta$ deformation, namely $\bar{\beta}_\pi\neq\bar{\beta}_\nu$.
For example, when $\chi_\nu = -\chi_\pi = -1.2$, the system results in
two degenerate quadrupole deformations:  $(\bar{\beta}_\pi,\bar{\beta}_\nu,\bar{\gamma}_\pi,\bar{\gamma}_\nu)=(0.73,~1.37,~0^\circ,~0^\circ)$ and $(1.37,~0.73,~60^\circ,~60^\circ)$.
Further analysis reveals
that the two generated minimal points described by $(\bar{\beta}_\rho,\bar{\gamma}_\rho)$
are connected through a minimal valley.
This valley is characterized by $\tilde{\beta}_\pi=\tilde{\beta}_\nu=\tilde{\beta}=\sqrt{\frac{7}{7+\chi_\mathrm{c}^2}}$ and $\tilde{\gamma}_\pi=\tilde{\gamma}_\nu=\tilde{\gamma}$ with $\tilde{\gamma}\in[0^\circ,60^\circ]$, yielding a $\gamma$-independent potential landscape with $V_{QQ}(\tilde{\beta}_\rho,\tilde{\gamma}_\rho)=-\frac{14}{14+\chi_\mathrm{c}^2}N_\pi N_\nu$. Note that the two terminal points of the minimal valley at $\tilde{\gamma}=0^\circ$ and $60^\circ$ are very close but do not coincide with the global minimal points described by $(\bar{\beta}_\pi, \bar{\beta}_\nu, \bar{\gamma}_\pi, \bar{\gamma}_\nu)$, except in the case where $\chi_\mathrm{c} = 0$. In this special case, the system at the mean-field level becomes fully $\gamma$-unstable, as evident from Eq.~(\ref{V}) with $\chi_\pi = \chi_\nu = 0$.
In general, the potential values within the minimal valley are slightly higher than the ground-state energy defined \ref{Eg}, with the energy gap (potential barrier height) given by
\begin{eqnarray}\label{deltaE}
\Delta E \equiv V_{QQ}(\tilde{\beta}_\rho, \tilde{\gamma}_\rho)-{V_{QQ}}_{\mathrm{min}}(\bar{\beta}_\rho, \bar{\gamma}_\rho)=\frac{\chi_{\mathrm{c}}^2}{14 + \chi_{\mathrm{c}}^2} N_\pi N_\nu\, .
\end{eqnarray}
This suggests a degree of $\gamma$-softness between the two degenerate global minimal points. When finite-$N$ effects are additionally considered, it may become challenging to distinguish between the case of two degenerate deformed minima connected by a shallow potential barrier and a $\gamma$-soft triaxial configuration in realistic systems. It should be mentioned that existence of two degenerate minima in this context differs fundamentally from the shape coexistence associated with crossing-shell excitations described by the IBM-2 with configuration mixing~\cite{Nomura2013,Nomura2017}. More accurate description of the potentials may require angular momentum projection on potential surfaces, which goes beyond the scope of this work.

In fact, the mean-field phase diagram of the IBM-2 encompassing various quadrupole shapes has been well established in prior studies~\cite{Arias2004,Caprio2004,Caprio2005}, based on an $F$-spin invariant Hamiltonian form~\cite{Iachellobook}. For instance, phase transitions leading to triaxial phases were investigated in \cite{Arias2004}, where a new critical point between spherical and triaxial shapes was reported; the rich phase structure of the IBM-2 and multiple order parameters in quantum phase transitions were revealed in \cite{Caprio2004}. A more comprehensive mean-field analysis of the IBM-2 phase diagram, along with quantal calculations, has been provided in \cite{Caprio2005}. In the $F$-spin invariant Hamiltonian, the like-boson quadrupole terms, $\hat{Q}_\rho \cdot \hat{Q}_\rho$ with $\rho = \pi, \nu$ are added to the Hamiltonian in addition to the currently adopted $\hat{Q}_\pi \cdot \hat{Q}_\nu$ term. The resulting SU(3)$_{\pi\nu}^\ast$-like triaxial configuration is described by $\hat{H}^\prime=-(\hat{Q}_\pi^{\chi_\pi}+\hat{Q}_\nu^{\chi_\nu})\cdot(\hat{Q}_\pi^{\chi_\pi}+\hat{Q}_\nu^{\chi_\nu})$, where $\chi_\pi=-\chi_\nu\simeq\pm\frac{\sqrt{7}}{2}$. The associated ground-state equilibrium in the $N_\pi = N_\nu \rightarrow \infty$ limit is characterized by $\bar{\beta}_\pi = \bar{\beta}_\nu$ and $\bar{\gamma}_{\pi} = 60^\circ - \bar{\gamma}_\nu \simeq 0^\circ$ or $60^\circ$~\cite{Dieperink1982,Arias2004,Caprio2004}. Note that the ideal SU(3)$_{\pi\nu}^\ast$ limit in the IBM-2 represents a globally triaxial deformation of two-fluid systems~\cite{Dieperink1982,Caprio2005}.
In this limit, the global triaxial deformation at the mean-field level is manifest by the equilibrium configuration, where a prolate deformed proton fluid coupled to an oblate deformed neutron fluid, and vice versa~\cite{Dieperink1982}. For example, using the coherent state defined in (\ref{coherent}), one can derive that the mean-field potential for $\hat{H}^\prime$ with $\chi_\nu = -\chi_\pi = -1.2$ in the large-$N$ limit has only one minimal point located at $(\bar{\beta}_\pi,~\bar{\beta}_\nu,~\bar{\gamma}_\pi,~\bar{\gamma}_\nu)\simeq(1.35,~1.35,~58.4^\circ,~1.6^\circ$). Clearly, this differs from the cases described by $\langle-\hat{Q}_\pi^{\chi_\pi} \cdot \hat{Q}_\nu^{\chi_\nu}\rangle$, which under the same parameters may generate two degenerate potential minima located at $\bar{\gamma}_\pi=\bar{\gamma}_\nu=0^\circ$ and $60^\circ$, respectively, as discussed above. However, the two degenerate deformed minima in the $\gamma$ direction are rather shallowed (see Eq.~(\ref{deltaE})), suggesting a $\gamma$-soft rotor structure in realistic situations. This scenario is expected to occur primarily when $\chi_\pi=-\chi_\nu\simeq\pm\frac{\sqrt{7}}{2}$, which aligns well with the spectral features
obtained from numerical calculations as exemplified in Fig.~\ref{F2}.

Contrastingly, when $\chi_\pi=\chi_\nu=\chi_\mathrm{c}$, if follows from Eq.~(\ref{V}) that the ground-state equilibrium deformation is characterized by $\bar{\beta}_\pi=\bar{\beta}_\nu=(\mid\chi_\mathrm{c}\mid+\sqrt{14+\chi_\mathrm{c}^2})/\sqrt{14}$ together with
$\gamma_\pi=\gamma_\nu=0^\circ$ or $60^\circ$, depending on whether $\chi_\mathrm{c}<0$ or $\chi_\mathrm{c}>0$. The resulting ground state energy can be expressed in a unified form as
\begin{eqnarray}\label{Eg1}
E_g^\prime&\equiv&V_{QQ_\mathrm{min}}(\bar{\beta}_\pi,\bar{\beta}_\nu,\bar{\gamma}_\pi,\bar{\gamma}_\nu)|_{\chi_\pi=\chi_\nu=\chi_\mathrm{c}}\\ \nonumber
&=&-N_\pi N_\nu\frac{\Big(\mid\chi_\mathrm{c}\mid+\sqrt{14+\chi_\mathrm{c}^2}\Big)^2}{7\Big(14+\chi_\mathrm{c}^2+\sqrt{\chi_\mathrm{c}^2(14+\chi_\mathrm{c}^2)}\Big)^2}\\ \nonumber
&~~&\times \Big[98+21\chi_\mathrm{c}^2+\chi_\mathrm{c}^4+\sqrt{\chi_\mathrm{c}^2(14+\chi_\mathrm{c}^2)^3}\Big]\, .
\end{eqnarray}
These results are similar to those obtained from the IBM-1 described by the quadrupole-quadrupole interaction without
distinguishing between protons and neutrons. Specifically, using the coherent state method~\cite{VC1981}, the corresponding mean-field potential function in the IBM-1 is given by~\cite{Iachello2004}
\begin{eqnarray}\label{VQQ1}
V_{QQ}^{\prime\prime}(\beta,\gamma)&\equiv&\langle-\hat{Q}^\chi\cdot\hat{Q}^\chi\rangle\\ \nonumber
&=&-\frac{N}{1+\beta^2}\Big[5+(1+\chi^2)\beta^2 \Big]-\frac{2N(N-1)\beta^2}{7(1+\beta^2)^2}\\ \nonumber
&~&\times\Big[14-2\sqrt{14}\chi\beta \mathrm{cos}(3\gamma)+\chi^2\beta^2 \Big]\, ,
\end{eqnarray}
where $\hat{Q}^{\chi}=(s^\dag\times\tilde{d}+d^\dag\times\tilde{s})^{(2)}+\chi(d^\dag\times\tilde{d})^{(2)}$.
In the large-$N$ limit, the linear $N$-dependent term in (\ref{VQQ1}) can be ignored, leaving only the second term proportioning to $N^2$.
Then, one can determine the optimal values of the deformation parameters $\beta$ and $\gamma$, which are given by $\bar{\beta}=(\mid\chi\mid+\sqrt{14+\chi^2})/\sqrt{14}$ together
with $\bar{\gamma}=0^\circ$ for $\chi<0$ or $\bar{\gamma}=60^\circ$ for $\chi>0$. Accordingly, the ground-state energy can be expressed as a function of $\chi$,
\begin{eqnarray}\label{Eg2}
E_g^{\prime\prime}&\equiv&V_{QQ_\mathrm{min}}(\bar{\beta},\bar{\gamma})|_{N\rightarrow\infty}\\ \nonumber
&=&-N^2\frac{\Big(\mid\chi\mid+\sqrt{14+\chi^2}\Big)^2}{7\Big(14+\chi^2+\sqrt{\chi^2(14+\chi^2)}\Big)^2}\\ \nonumber
&~~&\times \Big[98+21\chi^2+\chi^4+\sqrt{\chi^2(14+\chi^2)^3}\Big]\, .
\end{eqnarray}
Clearly, by setting $N_\pi=N_\nu=N$ and $\chi_\mathrm{c}=\chi$, Eq.~(\ref{Eg1}) may take the same form as Eq.~(\ref{Eg2}),
with the optimal values $\bar{\beta}_\pi=\bar{\beta}_\nu=\bar{\beta}$ and $\bar{\gamma}_\pi=\bar{\gamma}_\nu=\bar{\gamma}$.
This indicates that the two-fluid geometry described in this case is equivalent to that for one-fluid system at the mean-field level. In fact, if the quadrupole deformation parameters $\beta_\pi(\gamma_\pi)$ and $\beta_\nu(\gamma_\nu)$ are assumed to take identical values, the two-fluid potential function in Eq.~(\ref{V}) at $\chi_\pi=\chi_\nu$ reduces directly to the leading order term of the one-fluid functional given in Eq.~(\ref{VQQ1}), up to a scale factor.
One can thus obtain the well-known scenarios resulting from $\chi_\pi=\chi_\nu=\chi_\mathrm{c}$, including SU(3)-like prolate deformation for $\chi_\mathrm{c}<0$~\cite{Giannatiempo2012}, $\overline{\mathrm{SU(3)}}$-like oblate deformation for $\chi_\mathrm{c}>0$, and O(6)-like $\gamma$-unstable deformation for $\chi_\mathrm{c}=0$.
Consequently, the $\gamma$-soft deformation associated with two degenerate deformed minima achieved under the parameter condition $\chi_\pi=-\chi_\nu$ is not expected to occur in these cases.
In other words, collective modes corresponding to these familiar situations can only generate conventional
collective structures characterized by $B_{4/2}>1.0$. This, in turn, supports
the notion that $B(E2)$ anomaly phenomenon is more likely to occur in a nuclear system
where neutrons and protons exhibit different intrinsic quadrupole
deformations, such as the $\chi_\pi=-\chi_\nu$ cases discussed in the IBM-2.
These findings further confirm that the two-fluid boson model (IBM-2) can display much richer collective structures compared to the one-fluid case (IBM-1)~\cite{Iachellobook}.
Additionally, it is worth noting that a more general analysis of the correspondence between the IBM-2 and IBM-1 should adopt
the $F$-spin invariant IBM-2 Hamiltonian form~\cite{Arias2004,Caprio2004,Caprio2005}. It can be justified that, under the condition $N=N_\pi+N_\nu$ and $\chi_\pi=\chi_\nu$, both the two-fluid geometry and the low-lying dynamics (totally symmetric states) in the IBM-2 reduce exactly to those of the IBM-1~\cite{Iachellobook}.

\begin{center}
\vskip.2cm\textbf{(C). Descriptions of experimental data}
\end{center}\vskip.2cm

\begin{figure}
\begin{center}
\includegraphics[scale=0.25]{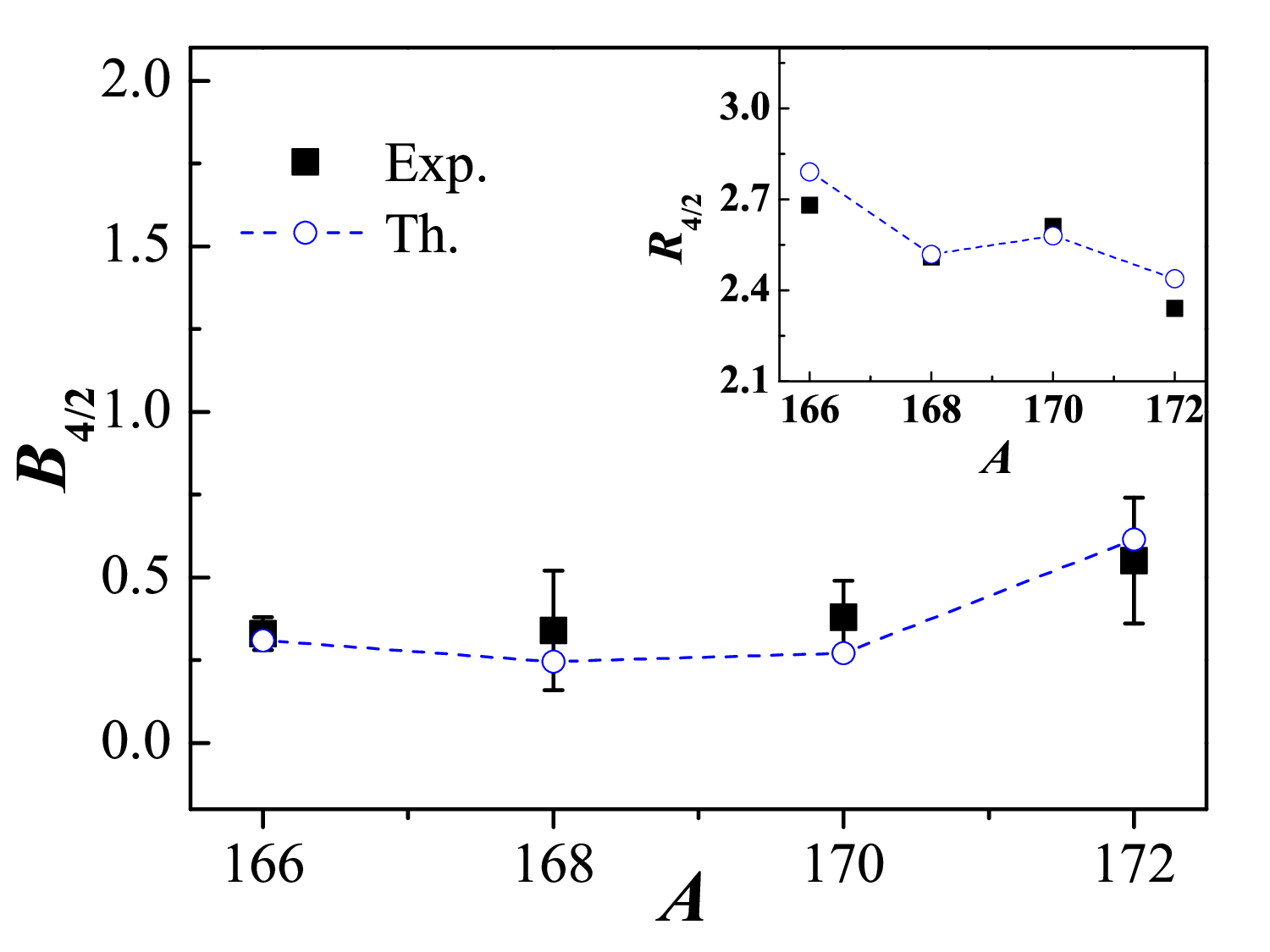}
\caption{The ratios of $R_{4/2}$ and $B_{4/2}$ in $^{166}$W~\cite{Saygi2017}, $^{168,170}$Os~\cite{Grahn2016,Goasduff2019}, and $^{172}$Pt~\cite{Cederwall2018} are shown as a function of the mass number $A$ to compare with the calculated results with the parameters in the consistent $Q$ Hamiltonian adopted as $\eta=0.036$ and $\chi_\pi=-\chi_\nu=1.18$.  }\label{F3}
\end{center}
\end{figure}

To examine the physical implications of the exotic $B(E2)$ features identified above, we compare the available data for the low-lying states in $^{172}$Pt, $^{170}$Os, $^{168}$Os, and $^{166}$W, all of which have been observed to exhibit anomalous $E2$ behaviors~\cite{Grahn2016,Saygi2017,Cederwall2018,Goasduff2019}, with the IBM-2 calculations using the consistent-$Q$ Hamiltonian defined in (\ref{H}).
To reduce the number of adjustable parameters and provide a qualitative explanation of the experimental data, $\eta$ and $\chi_{\pi(\nu)}$ have been treated as constants for all nuclei. Their values will completely determine (up to an overall scale factor) the dynamical structure of a two-fluid boson system with given $N_\pi$ and $N_\nu$. Additionally, since the proton configuration in these neutron-deficient nuclei is assumed to be hole-like and the neutron configuration is particle-like (relative to the $N=Z=82$ shell), we have assumed positive $\chi_\pi$ and negative $\chi_\nu$, with $\chi_\pi=-\chi_\nu$ in the calculations~\cite{Dieperink1982}.
With these constraints, the values of the $B_{4/2}$ and $R_{4/2}$ ratios solved from the IBM-2 are presented in Fig.\ref{F3} and compared with the available experimental data. As depicted in Fig. \ref{F3}, a remarkable consistency between experimental data and theoretical calculations can be clearly observed from the evolution of these ratios with the mass number $A$. In addition to the $B(E2)$ anomaly feature, where $B_{4/2} \ll 1.0$, the fluctuation in $R_{4/2}$ alongside nearly constant changes in $B_{4/2}$ are shown to be well reproduced by the IBM-2 calculations. Notably, $^{166}$W and $^{170}$Os are both associated with a total boson number of $N_\mathrm{B}=9$, while $^{168}$Os and $^{172}$Pt are associated with $N_\mathrm{B}=8$. Therefore, the differences in the $R_{4/2}$ values among these nuclei can primarily be attributed to the distinction in valence nucleon configurations, as all relevant parameters have been fixed as constants, as shown in Fig. \ref{F3}. This observation reflects the microscopic aspects of the IBM-2 in explaining the $B(E2)$ anomaly, especially in contrast to previous IBM-1 descriptions that utilized high-order interaction terms\cite{Zhang2022,Zhang2024,Pan2024,Wang2020,Zhang2025}.

\begin{table}
\caption{The calculated $B(E2)$ transitional rates (unit in W.u.) for the lowest-lying states are presented to compare with the available experimental data with "-"
denoting unknown data. In the calculations, the parameters in the Hamiltonian are taken as same as those adopted in Fig.~\ref{F3} and the effective charges $e_\pi=e_\nu$ in (\ref{TE2}) are determined from reproducing experimental value of $B(E2;2_1^+\rightarrow0_1^+)$ for each nucleus.}
\begin{center}
\label{T1}
\begin{tabular}{ccc|cc|cc|cc}\hline\hline
$L_i^\pi\rightarrow L_f^\pi$ &$^{172}$Pt&Th.&$^{170}$Os&Th.&$^{168}$Os&Th.&$^{166}$W&Th.\\ \hline
$2_1^+\rightarrow0_1^+$&49($11$)&49&97$_{-9}^{+9}$&97&74(13)&74&150(9)&150\\
$4_1^+\rightarrow2_1^+$&27($7$)&30&38$_{-7}^{+13}$&26&25(13)&18&50(7)&46\\
$6_1^+\rightarrow4_1^+$&-&49&-&86&-&51&18(4)&85\\
$8_1^+\rightarrow6_1^+$&-&46&-&74&-&38&-&56\\
$2_2^+\rightarrow0_1^+$&-&0.4&-&1.8&-&0.9&-&4.0\\
$0_2^+\rightarrow2_1^+$&-&16&-&7.1&-&5.0&-&0.03\\ \hline
$E(2_2^+)/E(2_1^+)$&-&1.45&-&1.37&-&1.27&-&1.26\\
$E(2_3^+)/E(2_1^+)$&-&2.67&-&2.60&-&2.52&-&2.65\\
$E(0_2^+)/E(2_1^+)$&-&1.72&-&1.56&-&1.37&-&1.22\\
\hline\hline
\end{tabular}
\end{center}
\end{table}

\begin{figure}
\begin{center}
\includegraphics[scale=0.32]{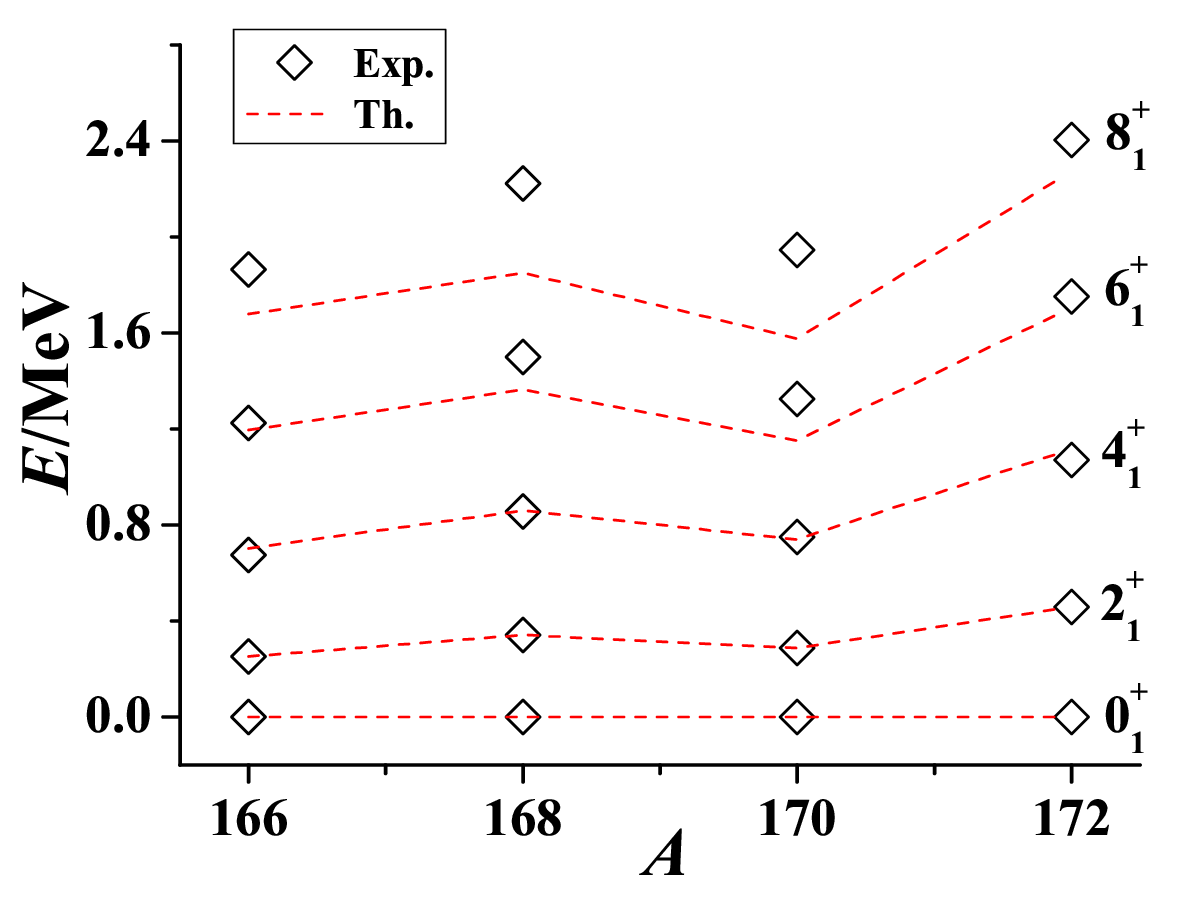}
\caption{The level energies of yrast states in $^{172}$Pt, $^{168,170}$Os and $^{166}$W are presented as a function of the mass number $A$ to compare with the model results. In the calculations, the parameters are taken as same as those adopted in Fig.~\ref{F3} except that an overall scale factor $\varepsilon_0$ in the Hamiltonian (\ref{H}) is additionally adjusted to reproduce the experimental $E(2_1^+)$ value for each nucleus.}\label{F4}
\end{center}
\end{figure}

More results for $B(E2)$ transitions are listed in Table~\ref{T1}, where it is evident that the available data can generally be well reproduced by the theoretical calculations. In particular, the results indicate that the anomalously depressed $B(E2)$ transitions in these neutron-deficient nuclei may occur to some extent for yrast states with $L > 4$. However, the $B(E2; L_1^+ \rightarrow (L-2)_1^+)$ transitions remain stronger than the inter-band $E2$ transitions, such as $B(E2; 2_2^+ \rightarrow 0_1^+)$ or $B(E2; 0_2^+ \rightarrow 2_1^+)$. As further observed in Table~\ref{T1}, one may find that there is an odd-even staggering in the calculated results for $B(E2; L_1^+ \rightarrow (L-2)_1^+)$ with $L=2m$ and $m=1,~2,~3,\cdots$. This is primarily attributed to the simple parameter assumptions $\chi_\pi=-\chi_\nu$ adopted in the consistent-$Q$ Hamiltonian (\ref{H}). The odd-even staggering feature could be partially mitigated by rearranging the parameters or adding additional interactional terms to the Hamiltonian, which should be pursued once more data becomes available. Additionally, it is shown that $B(E2; 6_1^+ \rightarrow 4_1^+)$ in $^{166}$W is evidently overestimated by the model calculations under the simple parameter assumptions of $\chi_\pi = -\chi_\nu$. Undoubtedly, the model descriptions can be further refined by adjusting the parameters individually for each nucleus. For example, resetting $\chi_\pi = 1.32$ and $\chi_\nu = -1.25$ for $^{166}$W may yield the $B(E2)$ resutls (unit in W.u.) with $B(E2; 2_1^+ \rightarrow 0_1^+) \approx 150$, $B(E2; 4_1^+ \rightarrow 2_1^+) \approx 62$, $B(E2; 6_1^+ \rightarrow 4_1^+) \approx 15$ and $B(E2; 8_1^+ \rightarrow 6_1^+) \approx 15$. Apart from the $B(E2)$ transitions, the typical energy ratios listed in Table~\ref{T1} suggest that non-yrast states, which have not yet been observed experimentally, may appear at low energy due to band mixing. These results are generally consistent with the previous IBM-1 analysis of the relevant nuclei~\cite{Zhang2022,Zhang2024,Pan2024,Wang2020,Zhang2025}.

To get further insights, we also examined the quadrupole moment of the $2^+$ states, which is defined by
\begin{eqnarray} Q(2^+)=\langle L^\pi,M=L\mid \hat{Q} \mid L^\pi,M=L\rangle|_{L^\pi=2^+}\, ,
\end{eqnarray} where $\hat{Q}=\sqrt{\frac{16\pi}{5}}T(E2)$ is given in Eq.~(\ref{TE2}).
Using the parameters given in Table~\ref{T1}, the calculations indicate $Q(2_g^+) = Q(2_1^+) > 0$ and $Q(2_\beta^+) = Q(2_3^+) < 0$ for all these neutron-deficient nuclei. However, when the the signs of $\chi_\pi$ and $\chi_\nu$ are simultaneously inverted, all the calculated results remain unchanged except for the quadrupole moments, which become $Q(2_g^+) < 0$ and $Q(2_\beta^+) = Q(2_3^+) > 0$. This indicates that the ground band and $\beta$ band in a $B(E2)$ anomaly system likely correspond to entirely distinct quadrupole deformations, providing additional references for future experimental measurements. Nonetheless, whether this prediction is robust against $\gamma$ softness requires further experimental validation.

Regarding low-lying levels, only those for yrast states are currently known from experiments, and the results as a function of the mass number $A$ are shown in Fig.~\ref{F4} for comparison with the theoretical calculations. As depicted in Fig.~\ref{F4}, the level evolutions are generally well described by the calculated results. However, some quantitative deviations may appear for levels with high spins. For example, $E(8_1^+)$ in $^{168,170}$Os is clearly underestimated by the theoretical calculations, although the overall trend of level evolution in experiments still follows the theoretical predictions closely. In general, such quantitative deviations in level energies can be improved by adding a rotational term $\hat{L}\cdot\hat{L}$ to the Hamiltonian in Eq. (\ref{H}). The modified consistent-$Q$ Hamiltonian has been successfully applied to describe the structural evolutions in the neutron-rich W, Os, and Pt isotopes with parameters derived from the microscopic Gogny energy density functional~\cite{Nomura2011II}. It is interesting to note that $\chi_\pi > 0$ and $\chi_\nu < 0$ can also be obtained for many neutron-rich nuclei using the constraint from microscopic mean-field calculations, as analyzed in \cite{Nomura2011II,Nomura2011III}. However, no $B(E2)$ anomaly phenomenon has been observed in the relevant nuclei. This confirms that the exotic modes with $B_{4/2} < 1.0$ identified here are indeed distinct from those well-known in the IBM-2~\cite{Iachellobook}. On the other hand, $\chi_\pi$ and $\chi_\nu$ in the quadrupole operators are suggested to have the same sign if valence protons and neutrons both have either hole or particle configurations, according to the assumptions proposed in \cite{Dieperink1982}. Consequently, the current analysis naturally leads to the conclusion that no $B(E2)$ anomaly is observed in neutron-rich triaxial nuclei, such as $^{190,192}$Os~\cite{Allmond2008}, which share the same boson numbers as their neutron-deficient partners, $^{168,170}$Os, discussed above.

Generally speaking, the results obtained from the present model are qualitatively similar to those derived from different IBM-1 Hamiltonian involving high-order terms, when concerning a given nucleus associated with $B(E2)$ anomaly~\cite{Zhang2022,Zhang2024,Pan2024}. Specifically, all models predict $B(E2)$ anomaly along the yrast lines and low-energy non-yrast states. Quantitative differences are primarily attributed to variations in parameter tuning across different models. For instance, the current calculations for $^{168}$Os predict an excitation energy of the $0_2^+$ state at $E(0_2^+)\approx0.43$ MeV and a weak $E2$ transition to the yrast state with $B(E2;0_2^+\rightarrow2_1^+)\approx5.0$ (in W.u.). These results are comparable to those denoted as IBM$_b$ in Table 1 of Ref.~\cite{Zhang2022} and those present in \cite{Zhang2024}, where the excitation energy $E(0_2^+)$ is calculated as $0.57$ MeV and $0.50$MeV, respectively, along with the $E2$ transition $B(E2;0_2^+\rightarrow2_1^+)\approx0.3$ (in W.u.) and $0.0001$ (in W.u.). Conversely, the results denoted as IBM$_a$ in Table 1 of Ref.~\cite{Zhang2022} indicate that imposing stricter constraints on parameters leads to a lower $0_2$ state with $E(0_2^+)\approx0.29$ MeV and thus a relative stronger interband transition $B(E2;0_2^+\rightarrow2_1^+)\approx18$ (in W.u.). Notably, the model parameters in these calculations have been adjusted to reproduce the experimental value $E(2_1^+)=0.34$ MeV, despite the adoption of different Hamiltonian. If relaxing this constraint on parameters, one can, in principle, obtain higher excitation energies for nonyrast states, as demonstrated in \cite{Pan2024}, where calculations based on parameters determined from a global fit yield $E(2_1^+)=0.54$MeV and $E(0_2^+)=1.35$MeV for $^{168}$Os. Similar situations are also observed in calculations for other relevant nuclei. All these differences may provide additional references for the future measurements.

\begin{center}
\vskip.2cm\textbf{IV. Summary}
\end{center}\vskip.2cm

In summary, the anomalous collective modes with $B_{4/2} < 1.0$ have been identified within the framework of the IBM-2 using a simple Hamiltonian form. It is shown that such exotic modes, associated with strong band mixing, can emerge in the IBM-2 systems that are dominated by the quadrupole-quadrupole interaction over a certain range of model parameters. This finding is applied to describe the yrast structures of four neutron-deficient nuclei exhibiting $B(E2)$ anomalies, which provides a simple theoretical explanation for the $B_{4/2} < 1.0$ puzzle observed in experiments. We emphasize that the present analysis supports the rotor-mode-induced band-mixing perspective~\cite{Zhang2025} on the $B(E2)$ anomaly obtained within the IBM-1 framework, which, however, necessitates introducing high-order terms in the Hamiltonian~\cite{Zhang2022,Zhang2024,Pan2024,Wang2020,Zhang2025}. Nonetheless, the IBM-2, which distinguishes between protons and neutrons, can provide an equal-quality theoretical description of the available data but using a simple Hamiltonian with significantly fewer parameters. This contributes to advancing the microscopic understanding of the $B(E2)$ anomaly phenomenon.

Apart from the even-even neutron-deficient nuclei discussed above, similar observations of $B(E2)$ anomalies have also been reported~\cite{Zhang2021} in adjacent odd-A species and even in light nuclei~\cite{Kintish2014, Tobin2014}. Extending the present analysis to odd-A systems~\cite{Teng2025} and to other microscopic model approaches~\cite{Vasileiou2024} would be of great interest and is anticipated. Related work is in progress.
\bigskip

\begin{acknowledgments}
Support from the National Natural Science Foundation of China (12375113,12175097) and from LSU through its Sponsored Research Rebate Program as well as the LSU Foundation's Distinguished Research Professorship Program is acknowledged.
\end{acknowledgments}

%\newpage

%\section*{References}

\end{document}